\documentclass[aps,nofootinbib,twocolumn,showpacs,superscriptaddress,amssymb]{revtex4}
\date{\today}
\usepackage{graphicx}
\begin{document}

\title{Universal decay law in charged-particle emission and
exotic cluster radioactivity}

\author{C. Qi}
\affiliation{School of Physics, and State Key Laboratory of Nuclear
Physics and Technology, Peking University, Beijing 100871, China}

\author{F.R. Xu}
\affiliation{School of Physics, and State Key Laboratory of Nuclear
Physics and Technology, Peking University, Beijing 100871, China}

\author{R.J. Liotta}
\affiliation{KTH, Alba Nova University Center, SE-10691 Stockholm,
Sweden}

\author{R. Wyss}
\affiliation{KTH, Alba Nova University Center, SE-10691 Stockholm,
Sweden}

\begin{abstract}
A linear universal decay formula is presented starting from the
microscopic mechanism of the charged-particle emission. It relates
the half-lives of monopole radioactive decays with the $Q$-values of
the outgoing particles as well as the masses and charges of the
nuclei involved in the decay. This relation is found to be a
generalization of the Geiger-Nuttall law in $\alpha$ radioactivity
and explains well all known cluster decays. Predictions on the most
likely emissions of various clusters are presented.
\end{abstract}


\maketitle

The first striking correlation between the half-lives of radioactive
decay processes and the $Q$-values of the emitted particle was found
in $\alpha$-decay systematics by Geiger and Nuttall~\cite{gn} as,
\begin{equation}\label{gn-o}
\log T_{1/2}=aQ_{\alpha}^{-1/2}+b,
\end{equation}
where $a$ and $b$ are constants. However, the Geiger-Nuttall law in
the form of Eq.~(\ref{gn-o}) has limited prediction power since the
coefficients $a$ and $b$ change for the decays of each isotopic
series~\cite{vs}. It may also change within a single isotopic chain
when magic numbers are crossed~\cite{Buck90}. Intensive works have
been done trying to generalize the Geiger-Nuttall law for a
universal description of all detected $\alpha$ decay
events~\cite{Brown92,Moll97,Royer00,Poe06,Poe02,Delion04}. Here we
present a truly universal formula valid for the radioactivity
of all clusters, including $\alpha$-particles. This
will allow us to search for new cluster decay modes and to
carry out a simple and model-independent study of the decay
properties of nuclei over the whole nuclear chart.

We thus observe that the $Q$-value dependence in Eq.~(\ref{gn-o}) is
a manifestation of the quantum penetration of the $\alpha$-cluster
through the Coulomb barrier. But this equation ignores the
probability that the $\alpha$-particle is formed on the nuclear
surface starting from its four constituent nucleons moving inside
the mother nucleus. This is the cause of the limitations of the
Geiger-Nuttall law mentioned above. In general the decay process,
ranging from proton to heavier cluster radioactive decays, can be
described by a two-step mechanism~\cite{Tho54}. In the first step
the formation of the particle and its motion on the daughter nuclear
surface is established. In macroscopic models the clusterization
process is described by effective quantities adjusted to reproduce
as many measured half-lives as possible. This procedure has shown to
be very fruitful, providing a guide to experimental searches. In the
second step the cluster, with the formation amplitude and
corresponding wave function thus determined, is assumed to penetrate
the centrifugal and  Coulomb
barriers~\cite{Buck90,Brown92,Moll97,Royer00,Poe06,Gam28}. This
second step is well understood since the pioneering work of Gamow.
It is in fact one of the pillars of the probability interpretation
of quantum mechanics~\cite{Gam28}. Its great importance in
radioactive decay studies lies in the fact that within a given
cluster the penetrability process is overwhelmingly dominant. This
explains the great success of macroscopic models in describing
radioactive decay.

We intend here to include the cluster formation probability as well
as the corresponding penetration through the Coulomb barrier. We
start from the R-matrix description of the cluster decay
process~\cite{Tho54,Lan58}. This is the basis of all microscopic
calculations of cluster decay \cite{Lovas98}. The corresponding
decay half-life is,
\begin{equation}\label{life}
T_{1/2}=\frac{\hbar\ln2}{\Gamma_c} \approx \frac{\ln2}{\nu} \left|
\frac{H_l^+(\chi,\rho)}{RF_c(R)} \right|^2,
\end{equation}
where $\nu$ is the outgoing velocity of the emitted particle which
carries an angular momentum $l$. $R$ is a distance around the
nuclear surface where the wave function describing the cluster in
the mother nucleus is matched with the outgoing cluster-daughter
wave function. For the distance $R$ we will take the standard value,
i.e., $R=R_0(A_d^{1/3}+A_c^{1/3})$ where $A_d$ and $A_c$ are the
mass numbers of the daughter and cluster nuclei, respectively.
$H^+_l$ is the Coulomb-Hankel function and its arguments are
standard, i.e., $\rho=\mu\nu R/\hbar$ and the Coulomb parameter is
$\chi = 2Z_cZ_de^2/\hbar\nu$ with $\mu$ being the reduced mass and
$Z_c$ and $Z_d$ the charge numbers of the cluster and daughter
nucleus, respectively. The quantity $F_c(R)$ is the formation
amplitude of the decaying cluster at distance $R$. The penetrability
is proportional to $|H_l^+(\chi,\rho)|^{-2}$. Eq.~(\ref{life}) is
valid for all clusters and for spherical as well as deformed cases.
The ratio $N_l =RF_c(R)/H_l^+(R)$, and therefore the half-life
itself, is independent of the radius $R$~\cite{Lovas98}.

In microscopic theories the formation amplitude is evaluated
starting from the single-particle degrees of freedom of the neutrons
and protons that eventually become the cluster. This is generally a
formidable task which requires advanced computing facilities as well
as suitable theoretical schemes to describe the clusterization
process. It is therefore not surprising that the first calculations
of absolute decay widths (which require a proper evaluation of the
formation amplitude) were performed after the appearance of the
shell model. These calculations had limited success due to the small
shell model spaces that could be included at that time. Only later,
with better computing facilities, the calculated half-lives started
to approach the corresponding experimental values. We will not deal
with microscopic theories here. For details and references on this
subject, including an historical background, see
Ref.~\cite{Lovas98}.

Our aim is to find few quantities that determine the half-life.
Expanding in these quantities we hope to be able to find, at the
lowest order of perturbation, an expression of the half-life which
is as simple as the Geiger-Nuttall law but valid in general, i.e.,
for all isotopic series as well as all type of clusters. With this
in mind we notice that the Coulomb-Hankel function can be well
approximated by an analytic formula, which for the $l=0$ channel
reads~\cite{fro57},
\begin{equation}
H^+_0(\chi,\rho) \approx (\cot
\beta)^{1/2}\exp\left[\chi(\beta-\sin\beta\cos\beta)\right],
\end{equation}
where the cluster $Q$-value is $Q_c=\mu \nu^2/2$ and
\begin{equation}\label{cosb}
\cos^2\beta=\frac{Q_c R}{e^2Z_cZ_d}.
\end{equation}
One sees that $\cos^2\beta$ would be a small quantity if $Z_cZ_d$ is
large, i.e., for heavy and superheavy systems. In this case one can
expand the last term in a power series of $\cos\beta$. By defining
the quantities $\chi' = Z_cZ_d\sqrt{A/Q_c}$ and $\rho' =
\sqrt{AZ_{c} Z_d(A_d^{1/3}+A_c^{1/3})}$ where $A=A_d A_c/(A_d+A_c)$,
one gets, after some simple algebra,
\begin{equation}\label{gn-1}
\log T_{1/2}=a\chi' + b\rho' + \log \left(\frac{\cot\beta\ln 2}{\nu
R^2|F_c(R)|^2}\right) + o(3),
\end{equation}
where $a = e^2\pi\sqrt{2m}/(\hbar\ln10)$ and
$b=-4e\sqrt{2mR_0}/(\hbar\ln10)$ are constants ($m$ is the nucleon
mass). The first two terms dominate the Coulomb penetration and
$o(3)$ corresponds to the remaining small terms. But still the
strong dependence of the half-life upon the formation probability in
the third term of Eq.~(\ref{gn-1}) has to be taken into account. It
is very difficult to make a microscopic calculation of the formation
amplitude $F_c(R)$. But we can extract it from the experimental
half-lives data by using Eq.~(\ref{life}), i.e.,
\begin{equation}
\log |RF_c(R)|=\frac{1}{2}\log \left[ \frac{\ln
2}{\nu}|H^+_0(\chi,\rho)|^2\right] - \frac{1}{2}\log T^{{\rm
Expt.}}_{1/2}.
\end{equation}
Taking $R_0=1.2$~fm we evaluated the function $\log |RF_c(R)|$
corresponding to $\alpha$ as well as heavier clusters. We thus found
that the formation probabilities of $\alpha$ decays are
located in the range $\log |RF_c(R)| = -1.5\sim -0.75$~fm$^{-1/2}$.
The stability of the $\alpha$ decay formation amplitude explains the
success of the Geiger-Nuttall and other empirical laws where
formation mechanism is not explicitly embedded. However, for all
observed cluster decays, ranging from $\alpha$ to the heavier
$^{34}$Si, the formation amplitude changes as much as eight orders
of magnitude.

Yet we found that Eq.~(\ref{gn-1}) can still
be written as a simple linear formula which properly takes into
account the strong dependence of the formation amplitude upon the
cluster as well as the mother nuclear structure to a first order
of approximation.
This we have
archived by exploiting the property that for a given cluster
$N_0\equiv RF_c(R)/H_0^+(\chi,\rho)$ does not depend upon $R$.
Proceeding as above one readily obtains the relation,
\begin{equation}\label{foramp}
\log \left|RF_c(R)\right|\approx \log \left|R'F_c(R')\right|+
\frac{2e\sqrt{2m}}{\hbar\ln10}\left(\sqrt{R_0'}-\sqrt{R_0}\right)\rho',
\end{equation}
where $R'=R'_0(A_d^{1/3}+A_c^{1/3})$ is a value of the radius that
differs from $R$. Since for a given cluster any nuclear structure
would be carried by the terms $RF_c(R)$ and $R'F_c(R')$ in exactly
the same fashion, Eq.~(\ref{foramp}) implies that the formation
amplitude is indeed linearly dependent upon $\rho'$.
Therefore one can write,
\begin{equation}\label{gn-2}
\log T_{1/2}=a\chi' + b\rho' + c.
\end{equation}
We emphasize here that the coefficient $b$ in this relation is
different from that of Eq.~(\ref{gn-1}). That is, the terms $ b\rho'+ c$,
which do not depend upon $Q_c$, have to include the effects
that induce the clusterization in the mother nucleus.
Moreover, we found that the term $\log \cot\beta/\nu$ in Eq.~(\ref{gn-1}) varies only slightly
for all the cases investigated below, from a minimum of 0.94 to a maximum of 1.2. The
effects induced by this variation,
as well as the higher order terms in Eq.~(\ref{gn-1}), are to be taken into account by
a proper choosing of the constants $a$, $b$ and $c$.

Eq. (\ref{gn-2}) holds for all cluster radioactivities. We will call
this relation the universal decay law (UDL). A straightforward
conclusion from the UDL is that $\log T_{1/2}$ depends linearly upon
$\chi'$ and $\rho'$. This to be valid should include the
Geiger-Nuttall law as a special case. One sees that this is indeed
the case since $\rho'$ remains constant for a given $\alpha$-decay
chain and $\chi'\propto Q_c^{-1/2}$. Below we will probe these
conclusions, and the approximations leading to them.

\begin{figure}
\includegraphics[scale=0.45]{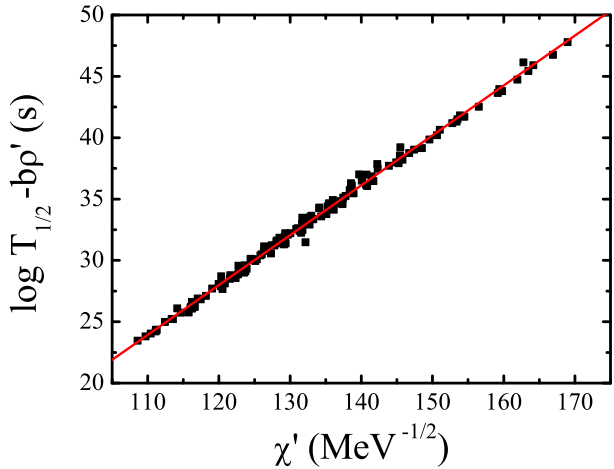}\\
\caption{(color online). UDL plots for the $\alpha$ decays of
even-even nuclei with $Z=78-118$. The straight line is given as
$a\chi'+c$.}\label{fig1}
\end{figure}

We will analyze g.s. to g.s. radioactive decays of even-even nuclei.
We select 139 $\alpha$ decay events from emitters with $78\leq
Z\leq108$ for which experimental data are available. We take the
data from the latest compilations of Refs.~\cite{Audi03,Audi03a} and
the lists of Refs.~\cite{zhang08,Pei07}. For the decay of heavier
clusters we have selected 11 measured events ranging from $^{14}$C
to $^{34}$Si~\cite{Bonetti07}. In order to perform the calculations
one has first to determine the values of the constants $a$, $b$ and
$c$. We carried out an extensive search of the best values for these
free parameters. Using a fitting procedure for the case of
$\alpha$-decay we obtained $a=0.4065$, $b=-0.4311$ and $c=-20.7889$.
The quality of the adjustment thus obtained can be seen in
Fig.~\ref{fig1}, where the values of  $\log T_{1/2}-b\rho'$ as a
function of  $\chi'$ is shown. The UDL reproduces the available
experimental half-lives within a factor of about 2.2. This compares
favorably with modern versions of the Geiger-Nuttall
law~\cite{Poe02}.

A significant deviation of the UDL in the Figure is the nucleus
$^{254}$Rf, at $\chi'=132.16~\text{MeV}^{-1/2}$, for which only the
lower limit of the half-life is available. This nucleus has the
value $T_{1/2}>1.5~$ms experimentally~\cite{Audi03}. The half-life
given by the UDL is $T_{1/2}=42$~ms, corresponding to a branching
ratio of $b=0.055\%$. A more precise measurement of this half-life
would be a welcome additional test of the UDL.

\begin{figure}
\includegraphics[scale=0.4]{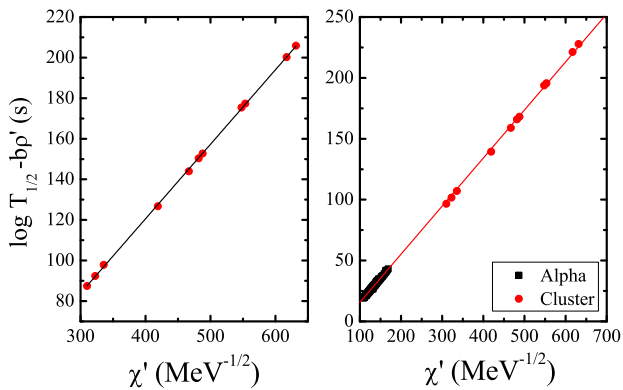}\\
\caption{(color online). Same as Fig.~\ref{fig1} but for the heavier
cluster decays (left panel) and both $\alpha$ and cluster decays
(right panel).}\label{fig2}
\end{figure}

We will now analyze cluster decay processes by comparing the
predictions of the UDL with the experimental data corresponding to
the decay of even-even nuclei mentioned above~\cite{Bonetti07}.
Using the parametrization set II in Table~\ref{table1} we plotted,
as before, the quantity $\log T_{1/2}-b\rho'$ as a function of
$\chi'$. As seen in the left part of Fig.~\ref{fig2} the agreement
between experiment and the UDL is excellent. The UDL reproduces the
available experimental half-lives within a factor of about 4.1.

Finally we consider all decays together, i.e., $\alpha$ as well as
heavier clusters. Using the parameter set III of Table~\ref{table1}
we obtained the results shown in the right panel of Fig.~\ref{fig2}.
Again the agreement between the UDL and experiment is excellent.

\begin{table}
\centering \caption{Coefficient sets of Eq.~(\ref{gn-2}) determined
by fitting to experimental data in $\alpha$ decay (I), cluster decay
(II) and both $\alpha$ and cluster decays (III), respectively. The
last column is given by the Coulomb barrier penetration term of
Eq.~(\ref{gn-1}) with $R_0=1.2$~fm. \label{table1}}
\begin{ruledtabular}
\begin{tabular}{ccccc}
& I($\alpha$) & II(cluster) & III($\alpha$+cluster) &IV \\
\hline
$a$ & ~~~0.4065 &~~~0.3671 & ~~~0.3949 &~0.4314\\
$b$ & ~~-0.4311 &  ~-0.3296 & ~~-0.3693 &-0.5015\\
$c$ &-20.7889 &-26.2681 & -23.7615 &\\
\end{tabular}
\end{ruledtabular}
\end{table}

Using the UDL it is straightforward to evaluate the half-lives of
all cluster emitters throughout the nuclear chart if reliable values
of the binding energies are provided. This we obtain by using the
latest compilation of nuclear masses~\cite{Audi03a}. With the
$Q$-values thus obtained we have evaluated the decay half-lives of
all isotopes included in that compilation by applying the UDL. We
thus found that in all cases the experimental values lie between the
ones calculated by using the parameters of the sets I and III in
Table~\ref{table1}, confirming the prediction power of the UDL. We
also found that nuclei favoring cluster decays are mostly located in
the trans-lead region.

In Table~\ref{table1} we also give the values of the coefficients $a$
and $b$ as provided by Eq.~(\ref{gn-1}). It can be seen that these values
are close to the corresponding fitted values, confirming that effects induced
by  $\log \cot\beta/\nu$ and higher-order terms in Eq.~(\ref{gn-1}) are small.

In summary, we have presented in this paper a simple formula that
provides with great precision the half-lives corresponding to
cluster decay. The formula is valid for all kind of clusters and for
all isotopic series, as expected since we derived it from the
general description of the decay half-life. This formula is of a
universal validity and therefore we call it universal decay law
(UDL). There are a few exceptions to this feature, in particular the
alpha-decay of $^{254}$Rf for which only the lower limit of the
half-life is available. The UDL predicts that this half-life should
be $T_{1/2}=42$~ms. A measurement of this number, as well as other
cases presented in this paper for heavy and superheavy nuclei which
may be of interest in present experimental facilities, would be most
welcome  to probe the extension of validity of the UDL. This law may
also help in the ongoing search of new cluster decay modes from
superheavy nuclei.

\section*{Acknowledgments}

This work has been supported by the Chinese Major State Basic
Research Development Program under Grant 2007CB815000; the National
Natural Science Foundation of China under Grant Nos. 10525520,
10735010 and 10875172 and the Swedish Science Research Council (VR).

\end{document}